\documentclass[12pt]{revtex4-1}
\usepackage{graphicx}
\usepackage{amssymb} 
\usepackage{amsmath}    
\usepackage{latexsym}   
\usepackage{bm}
\usepackage{times}
\usepackage{centernot}
\usepackage{float}
\usepackage{color}
\usepackage{wrapfig}
\usepackage{mathtools} 

\def\abr{\mathrel{\stackrel{\longrightarrow}{A\!B}}\!}
\def\bar{\mathrel{\stackrel{\longrightarrow}{B\!A}}\!}

\newtheorem{prop}{Proposition}

\begin{document}
\title{Synchronization of symbols as the construction of times and places}

\author{John M. Myers}
\email{myers@seas.harvard.edu}
\affiliation{Harvard School of Engineering and Applied Sciences, Cambridge, MA 02138, USA.}
\author{F. Hadi Madjid}
\email{gailmadjid@comcast.net}
  \affiliation{82 Powers Road, Concord, MA 01742, USA.}


\begin{abstract}
We demonstrate an unsuspected freedom in physics, by showing an
essential unpredictability in the relation between the behavior of clocks on
the workbench and explanations of that behavior written in symbols on the
blackboard. In theory, time and space are defined by clocks synchronized as
specified by relations among clock readings at the transmission
and reception of light signals; however spacetime curvature implies
obstacles to this synchronization.  Recognizing the need to handle bits and
other symbols in both theory and experiment, we offer a novel theory of
symbol handling, centered on a kind of ``logical synchronization,'' distinct
from the synchronization defined by Einstein in special relativity.

We present three \vspace*{-6pt}
things:
\begin{enumerate}
\item We show a need in physics, stemming from general relativity, 
  for physicists to make choices about what clocks to synchronize with what
  other clocks.
\item To exploit the capacity to make choices of synchronization, we provide a
  theory in which to express timing relations between transmitted symbols and
  the clock readings of the agent that receives them, without relying on any
  global concept of ``time''. Dispensing with a global time variable is a
  marked departure from current practice.
\item The recognition of unpredictability calls for more attention to
  behavior on the workbench of experiment relative to what can be predicted
  on the blackboard.  As a prime example, we report on the ``horse race''
  situation of an agent measuring the order of arrival of two symbols, to
  show how order determinations depart from any possible assignment of values
  of a time variable.
\end{enumerate}
\end{abstract}

  \maketitle

\section{Introduction}
This work began with a vision of an essential unpredictability in
physics, beyond quantum uncertainty.  A milestone in our pursuit of this
vision was our 2005 proof that in quantum physics, whatever evidence may be
on hand leaves open choices of explanations of that evidence.  That proof
depends only on the rule (the Born trace rule) that expresses evidence as
probabilities of outcomes and expresses explanations by density operators
paired with measurement operators.  Thus, resting only on the basic structure
of quantum theory, we find that the writing of a density operator or a wave
function to explain evidence involves an element of choice undetermined by
any application of logic to measured data.  As discussed in
\cite{05aop,16aop,18incomplete}, the proof is made possible by noticing that
physicists communicate in symbols, such as the numerals and the letters of
the alphabet on the page before you.  The proof and the recognition of the
dependence of physics on commuicating sequences of symbols makes an
opening in physics for agents and their symbols.  The opening demands
departing from the prevalent habit of theorists by respecting experimental
behavior on the workbench as conceptually distinct from any of its
theoretical expressions on the blackboard, leading to unexpected freedom in
the construction of times and places.
 
The next step in pursuing the vision is to pose the right question.  Our
first try was:
\begin{quote}
  How does physics change if we recognize that its equations are written by
  people who have choices undetermined by physical evidence?
\end{quote}
This question presents a stumbling block: the equations, once
written, look the same, regardless whether they reflect a person's free
choice or not.  We find a more promising question as:
\begin{quote}
  Does essential unpredictability show up in material behavior, for example in
  the behavior of clocks, in a way that warrants theoretical attention?
\end{quote}
To face the question, we have go back to basics.  Scientists try to
picture how the world works.  They build on what has gone before.  Their
experiments and theories co-evolve in a context of hitherto unappreciated
unpredictability.

The history of science over the past centuries cycles
between unifications and fragmentation.  Here we discuss how several
fragments can now jell into an unexpected unity, based on recognizing that:
(1) laws of physics do not write themselves, but are products of an evolving
species of organisms, namely people, and (2) discrepancies among clocks as
devices on the work bench call for more extensive description than can be
expressed by ``uncertainty''.   Together these mean that the theories and
experimental procedures promoted by physicists contend among themselves for a
place in cultural evolution.  Support for this point of view will emerge as we proceed.

The great unification in physics came half a millennium ago, with the
invention of the pendulum clock and the telescope on the experimental side
and of the mathematical formulation of derivatives and integrals---the
calculus---of Newton and Leibniz, on the theoretical side.  A great
unification sometimes requires, however, the {\it in}attention to certain
realities that might  embarrass it.  Newton saw the diversity of rhythms that
can be compared with one another not as the rich source of physics that we
shall find it to be, but as an embarrassment to be resolved by leading his
readers away from the diverse and independent pendulums they can see into an
abstraction (in Newton' words):
\begin{quote}
  Although time, space, place, and motion are very familiar to everyone, it
  must be noted that these quantities are popularly conceived solely with
  reference to the objects of the sense perception. And this is the source of
  certain preconceptions; to eliminate them it is useful to distinguish these
  quantities into absolute and relative, true and apparent, mathematical and
  common.

  1. Absolute, true, and mathematical time, in and  of itself and of
its own nature, without reference to anything external, flows uniformly and
by another name is called duration.  Relative, apparent, and common time is
any sensible and external measure (exact or nonuniform) of duration by means
of motion; such a measure---for example, an hour, a day, a month, a year---is
commonly used instead of true time.  \cite{newton}
\end{quote}

We take objection to Newton's picture, and are by no means the first to
object to it.  We aim to respect clocks on the work bench as having
contributions to make distinct from what can be asked of mathematical
formulations of ``time''; indeed we view the purpose of clocks as ``telling
time'' as a secondary purpose, by no means a defining purpose.

Einstein broke away from Newton's concepts of time and of space, but kept
more of them than one might think.  Although Einstein made the special
relativistic concept of time depend on ``clocks,'' these are not ``clocks on
the workbench'' but {\it proper clocks}, which are just as mathematical and
abstract as Newton's mathematical time.  The special-relativistic ``time''
defined by the use of (idealized) light signals to synchronize proper clocks
was relative to a choice of frame (thus ``relativised'') but this ``time''
inherits from the mathematical tradition of Newton the suppression of the
diversity of rhythms on the workbench.

That would not matter if all the bench rhythms could all be related to some
standard rhythm in any simple, ``objective'' way, but that is not the case.
The best we have are the time broadcasts supplied via the Global Navigation
Systems, the internet and cell phones.  Time broadcasts involve national metrology institutes (NMI's). There is not
single clock for for the world's time; each NMI has several clocks, and these
drift apart, so that the NMI nudges the clock rates to keep them from
excessive drift, both within an individual NMI and in the relations between
an NMI and clock reading transmitted to one NMI from another NMI. This is no
``objective'' business but a matter of intense negotiation.  As one of the
experts puts it:
\begin{quote} 
The fact is that time as we now generate it is dependent upon defined
origins, a defined resonance in the cesium atom, interrogating
electronics, induced biases, timescale algorithms, and random
perturbations from the ideal. Hence, at a significant level, time---as
man generates it by the best means available to him---is an
artifact. Corollaries to this are that every clock disagrees with
every other clock essentially always, and no clock keeps ideal or
``true'' time in an abstract sense except as we may choose to define
it. \cite{allan87}
\end{quote}

So much for the gap between relativity theory and the implementation of time
broadcasts.  Returning to general relativity, the very simplification of
assuming proper clocks leads to a shocking consequence.  Einstein's theory of
proper clocks, once he extended it into curved spacetime, challenges
its own conception of ``time'' in a way that, curiously, supports recognizing
choice and agency in the organization of even mathematically expressed
proper clocks.  But why should non-experts care about irregularities that
generally affect ``time on Earth'' only through its partition into time zones and in
the technology of the Global Navigation Systems that are adjusted on the scale
of nanoseconds to accommodate spacetime curvature?  Even for non-experts,
appreciating the collision of today's physical theory of time with itself
presents an opportunity think about some interesting situations free of
prevalent conceptual errors. Sec.\ \ref{sec:2}, backed up by the Appendix,
tells this story.

In the century after Einstein's theories of clocks in special and general
relativity came the digital communications revolution, along with startling
improvements in physical clocks, first atomic clocks and more recently
optical atomic clocks.  But a curious invariance showed up along the stages
of improvements in clock precision: getting any two clocks to tick as close
together as their evolving technology allows, requires steering their clock
rates.  Arranging for any pair of clocks to agree as closely as possible
continues to require, in effect, that agents adjust the ``pendulums of their
clocks'' in response to unpredictable discrepancies between the clocks.

The business of the computer networks and other digital networks that pervade
modern scientific life is to manipulate and communicate symbols.  Symbols
include numerals that convey readings of clocks that step computers through
their cycles of operation.  The clock of a computer in not primarily needed
to ``tell time'' but to tell the computer when to step.  A main purpose of
this paper is to introduce the concept of a clock as a tool of a
symbol-handling agent, primarily a tool for managing relations between chosen
rhythms in the face of unpredictable effects, and only secondarily as a tool
for ``telling time''.


While viewing clocks as tools of agents has potential advantages, realizing
that potential requires a novel conceptual framework in which to think about
and measure one rhythm in relation to another, without assuming any globally
available ``time.'' In earlier work we introduced a mathematical form for
expressing relations among biological rhythms \cite{JCS}.  In
Sec.\ \ref{sec:3} we repeat, with only minor changes, this mathematical form,
no longer confined to applications in biology, to make it available for its
relevance to fundamental and applied physics.  The basic ideas of the
mathematical form are: (1) the notion of an agent that handles symbols
sequentially, one after another, and records the symbols handled on what we
call a {\it clock tape}, and (2) the relation, called a {\it transmission
  relation} to express how symbols received by an agent fit into the sequence
on the agent's clock tape, as illustrated in Fig.\ \ref{fig:1}(a). For
present theoretical purposes, a network is a set of agents linked by
transmission relations.  This theoretical formulation is applicable to
networks in a large variety of situations and levels of detail of
description, from the world wide web to biochemical networks within a
bacterium.  A major novelty introduced in this paper is the means to express
communication networks without reference to any globally available ``time
coordinate''.  Transmission relations serve to express the timing aspect of
communications among agents of a network, without requiring any particular
assumption of how symbols are propagated; neither a metric nor indeed any
spacetime manifold need be assumed.  These transmission relations offer a
conceptual foundation for constructing ``times'' and ``places''.

The freedom to explore the construction of times and places stands in marked
contrast to prevalent designs for synchronizing digital sensor
networks---designs that approximately implement ``time'' as defined in
special relativity. An example of prevalent designs is the synchronization of
the global network of eight radio telescopes that produced the recent picture
of a black hole \cite{19EHT}.  Another example is the synchronization of
dispersed detecting devices in the Compact Muon Solenoid
(CMS) \cite{07synchMST}.  A third example is an undersea network of sensors
for which GPS signals are unavailable, in which synchronization is
implemented using the Precision Time Protocol (IEEE 1588) \cite{12synchMST}.
A more recent example is in \cite{19synchMST}.  We make no claim to improve
the efficiency by which synchronization is managed in these designs.  Rather,
we offer an alternative approach to synchronization, previously
unappreciated, that opens up novel avenues to investigation. The avenues we
have thought of so far center on transmission relations involved in
biological organisms, where different rhythms come into and drop out of
synchronization with other such rhythms, as discussed in \cite{JCS}.

Computer to computer communications offer another lesson from the workbench
that warrants theoretical attention: communicating digits from one computer
to another requires relations among clocks quite distinct from those
defined in special relativity, having to do with phasing of digit arrivals relative
to the clock that steps a receiving computer. Sec.\ \ref{sec:4} describes the
need for agents to adjust the rates of their clocks so that symbols arrive at
an agent during a suitable phase of the agent's clock, the condition of {\it
  logical synchronization}.  Maintaining logical synchronization requires
that agents respond to timing gradations beyond the reach of machinery used
to recognize distinct symbols, a finding well known to engineers of digital
hardware, but deserving more attention in theoretical physics.

Sec.\ \ref{sec:5} deals with what might be called the ``the machinery of
logic in motion.'' Critical to machinery for symbol manipulation is a tiny
balancing device called a {\it flip-flop}.  Occurring by the millions on the
silicon chips of digital systems, the flip-flop stores an elementary
symbol---a single bit.  The flip-flop works like a hinge that, flipped one
way, shows a 1, or if flopped the other way, shows a 0.  We discuss the
flip-flop as a balancing device that not only holds an elementary logical
value, but moves in response to changes in that value.  In a computer, a
flip-flop decides on the temporal order of a clock tick and a symbol arrival:
did the symbol arrive before or after the tick?  In a close race between the
clock tick and the symbol, the flip-flop can be tipped into an unstable
equilibrium, a condition that leads to {\it logical confusion} more complex
than anything expressed by ``measurement uncertainty.''  The experimental
demonstration of this logical confusion, illustrated by Figs.\ \ref{fig:osc}
and \ref{fig:glitch1}, prompts us to see clocks and their management as a topic on
its own, separable from what now strikes us as the problematical concept of
``global time.''  Concluding remarks occupy Sec.\ \ref{sec:6}.

\section{Agency and the theory of time and length}\label{sec:2}
Much of this paper, especially Sec. \ref{sec:4}, is concerned with
introducing the concept of {\it logical synchronization}, but current theoretical
physics hinges on a quite different form of synchronization, defined by
Einstein in special relativity, that we refer to as {\it Einstein
synchronization}.  Here we show how, in the curved spacetime of general
relativity, Einstein synchronization encounters obstacles, in a way that
makes an opening in theoretical physics for agents that make choices beyond
the reach of logic.  In this section we sketch the story, relegating its
justification to the Appendix.

From Einstein, theoretical physics inherits not just one but two theories of
time, space, and spacetime.  Special relativity postulates inertial frames as
free of acceleration and of gravitational influences.  Then `time' and
`length' are elegantly defined, relative to a choice of inertial frame, in
terms of the Einstein synchronization of proper clocks.  Einstein
synchronization is defined as a condition on readings of proper clocks when
they transmit and receive (theoretical) light signals.  One can picture the
time coordinate relative to an inertial frame as made available by an
infinitely fine, three-dimensional grid of Einstein-synchronized proper
clocks, so that every event coincides with a unique tick of a unique clock of
the grid.

The special-relativistic definitions of time and length in terms of
Einstein-synchronized proper clocks are the theoretical basis of the units of
measurement for time and length in the International System (SI).  But
corrections are needed.  To deal with acceleration and gravitation, Einstein
made special relativity hold only in vanishingly small spacetime regions
of globally curved spacetime.  The curvature of spacetime proved not just to
be theoretically attractive, for example in astrophysics, but modern
navigation systems, such as the Global Positioning System (GPS), depend on
the theory of light signals and clocks articulated in the theory of general
relativity.

In the theory of curved spacetime, there can be no inertial frame and no
infinitely fine grid of Einstein-synchronized proper clocks by which to
define time and length.  One still has the notion, discussed in the Appendix,
of {\it observer fields}, any of which is a set of not-necessarily proper
clocks so that every event coincides with a unique reading of a unique clock
of the observer field.  The notion of an observer field allows for
theoretical clocks that are {\it improper} in the sense of generating
readings at a rate that varies relative to co-present proper clocks.  But even
allowing for an observer field of improper clocks, curvature presents an
obstacle to having the clocks of an observer field be Einstein synchronized
with one another.  An observer field can be chosen such that subsets of a few
of its clocks can be Einstein-synchronized with one another, but that choice
precludes other choices that would Einstein synchronize other small sets of
clocks.

\begin{prop}\label{prop:2_4}
  For a generic curved spacetime, Einstein synchronization can be achieved,
  even with clock adjustment allowed, only for selected pairs of clocks; that
  is, the selection of some pairs of clocks to be synchronized excludes
  Einstein synchronization among other pairs of clocks.
\end{prop}

The requirement to select which clocks to Einstein synchronize with which
other clocks raises the question of who or what does the selecting, leading
us to the notion of an {\it agent}.  That requirement is also a hint that
times are necessarily local times, where by {\it local} we mean dependent on
choices made by agents.

\section{Symbol-handling agents}\label{sec:3}
In this section we offer a theory of symbol handling by which to express
relations among symbols communicated among clock-using agents, relations that
constitute a system of times and places adapted to their communication.  The
type of synchronization required for agents to communicate is the topic of
the following Sec.\ \ref{sec:4}.

As the term is used here, an {\it agent} has a ''local clock'' consisting of
a cyclic motion, e.g. a swinging pendulum that the agent can adjust, along
with the means to count cycles.  The count is a ``local time''.  Only in
special cases, however, is the clock of one agent Einstein-synchronized to
the clock of any other agent.  Thus, in general, the agent has available no
`time' as defined in special relativity.  In step with the ticks of its
clock, the agent deals with symbols sequentially.  We consider agents that
communicate symbols among themselves, as well as to and from an environment,
in rhythms set by their (adjustable) clocks.  For such agents we offer a
mathematical framework for expressing: (1) the record of the sequence of
symbols that an agent has dealt with; and (2) the timing of symbol exchange
among agents.  Agents linked in a communications network can work at very
different clock rates, and the framework offered needs no assumption of a
global time coordinate, nor of spacetime.

As its adjustable clock ticks, an agent executes moves, one after another,
each move involving a symbol.  The adjustable clock drives a tape, which we
call a {\it clock tape}, reminiscent of the tape of a Turing machine
\cite{turing}.  (We drop the assumption, made in our prior work
\cite{16aop}, that agents have the capability of a universal Turing machine.)
If one does think of a Turing machine with its infinite tape, then the clock
tape is an additional ``write-only'' tape.  The agent, as we now think of it,
has a memory, separate from the clock tape, that holds strings of symbols, and
the agent's action can depend on symbols held in its memory.  The symbol that
an agent records on a square of its clock tape at a move might be read from
its memory, written into its memory, received from another agent, transmitted
to another agent, or emerge from contact with an unknown realm (which we
associate with acts of guesswork \cite{16aop}, but will not discuss further
here).

Like the tape of a Turing machine, the clock tape is pictured as marked off
in squares, with only one square immediately visible to the agent at any
move.  As its clock ticks, the agent's clock tape advances by one square,
always in the same direction. (Unlike the tape of a Turing machine, the clock
tape is {\it not} erasable.)  By recording one symbol after another on the
squares of the clock tape an agent converts its temporal sequence of symbols
into a spatial sequence, like a film strip, amenable to mathematical
expression.

\subsection{Transmission relations}
Based on the above picture, we offer a mathematically expressed theory of
agents transmitting and receiving symbols.  Applications of this theory to
some simple, engineered digital networks is transparent, while other applications,
for example to cases of symbol transmission in biology, call for making
assumptions tailored to the case.

\begin{figure}[h!]
 \includegraphics[height=4 in]{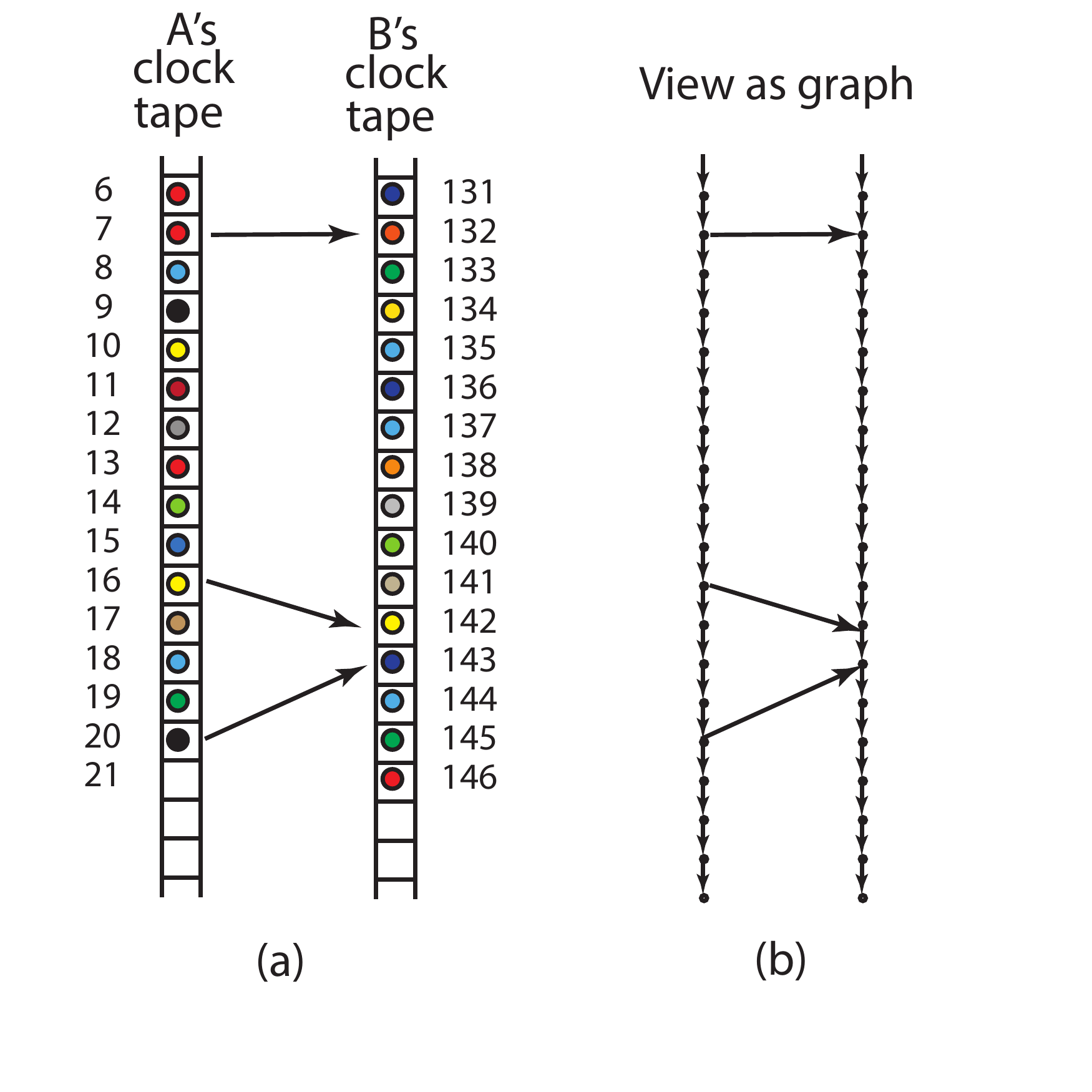}\vspace*{-.5 in}
 \caption{(a) Transmission from $A$ to $B$ from \cite{JCS,19spie}
   ; (b) Graph showing transmission relation from\cite{JCS,19spie} .\label{fig:1}}
  \end{figure}

We express the timing of the transmission of symbols from agent $A$ to an
agent $B$ by a {\it transmission relation} comprised of ordered pairs, each
pair consisting the square on $A$'s clock tape that records the transmitted
symbol and the square of $B$'s clock tape that records the reception of the
symbol transmitted by $A$, as illustrated in Fig.\ \ref{fig:1}(a).  Each
arrow from $A$ to $B$ indicates an ordered pair.  We can label successive
squares of $A$'s clock tape by successive integers and do the same for $B$'s
clock tape, so that integers serve as names for squares.  For example the
arrow from square 7 of $A$'s clock tape to square 132 of $B$'s clock tape
expresses the ordered pair (7,132), indicating that a symbol on square 7 of
$A$'s clock tape was transmitted to $B$ and recorded as received on square
132 of $B$'s clock tape.  This labeling by integers is ``local'' in that one
integer larger than another on $A$'s tape means one square recorded later
than another, but an integer larger on an $A$-tape than an integer on a
$B$-tape says nothing about temporal order of the squares on those two
distinct tapes.  Cross-tape temporal order is expressed {\it only} by
transmission relations.

Transmission relations that link the sequence of squares of the clock tape of
one agent to the sequence of squares of the clock tape of another agent are a
basic unit of analysis for the timing aspect of symbol handling.  It is the
clock tape that makes it possible to relate the rhythm of one agent to that
of another agent, without assuming any global time variable, thereby opening
up what we like to call ``two-clock physics''.
A transmission relation is indifferent to which symbol rests on which square
of a clock tape; thus it can be expressed by a graph, as illustrated in
Fig.\ \ref{fig:1}(b).
    
Now we generalize the transmission relation by allowing the
transmission relation to express influences weaker than literal transfer of a
symbol:
\begin{quote}
  The transmission relation from $A$ to $B$ links a symbol on $A$'s clock
  tape to a (possibly other) symbol on $B$'s clock tape if the symbol from
  $A$ directly influences the writing of a symbol on $B$'s clock tape.
\end{quote}
We emphasize again that transmission relations need no assumption of any
spacetime manifold.  One is free to make whatever assumptions one wants to
explain how a symbol moves from one agent to another agent, conditioned
of course by traditions, notably the tradition of treating the speed of light
as an absolute limit to the speed of propagation.

\subsection{Specialized properties of transmission relations}
\begin{figure}[h]
  \hspace*{1 in}
    \includegraphics[height=2.5 in]{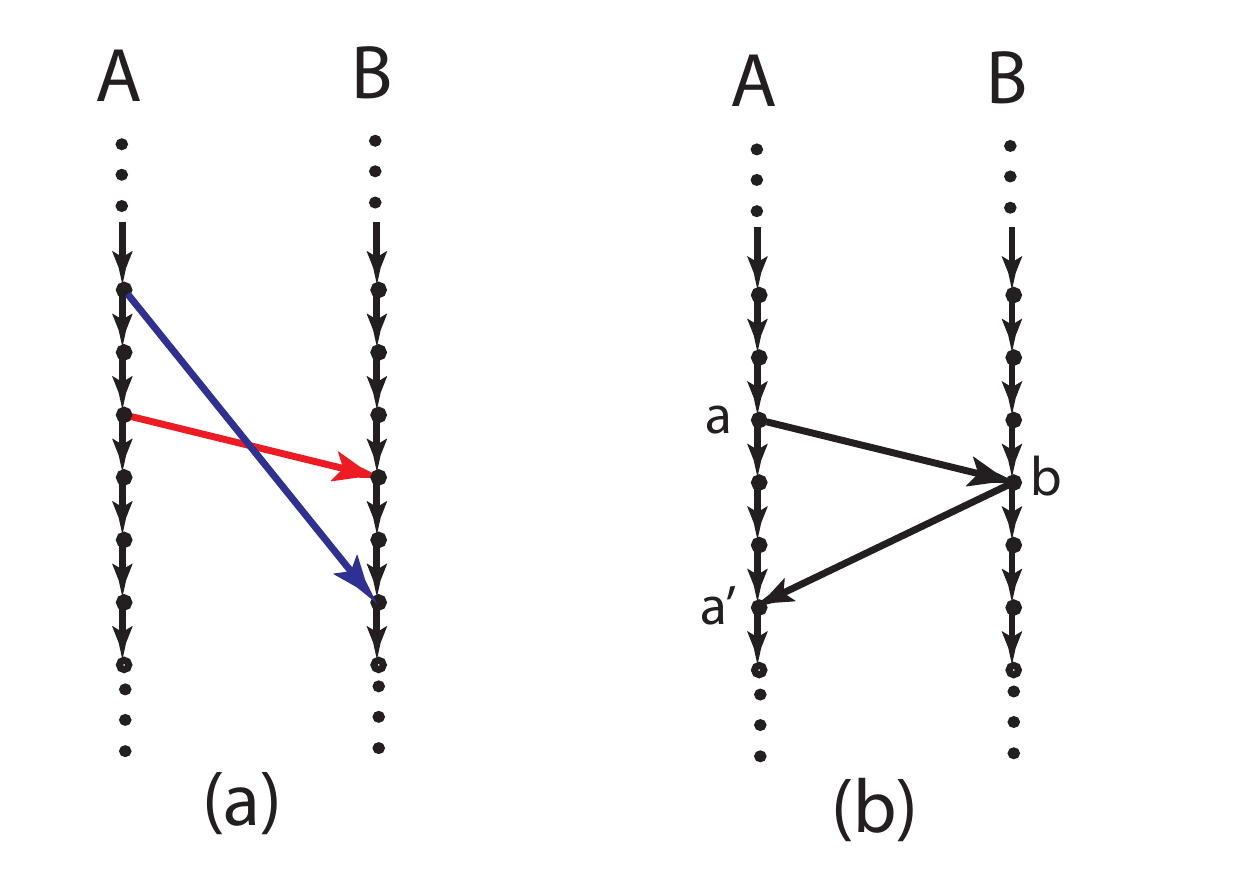}
    \caption{(a) Fragment of a transmission relation showing non-preservation
      of order from \cite{JCS} ; (b) example of $b \in B$ with an echo count
      relative to $A$ of 3 from \cite{JCS} . \label{fig:X1}}
 \end{figure}

We let $\abr$ denote a transmission relation from $A$ to $B$.  According to
the application, one or another property from the following list can be of
interest:
\begin{enumerate}
\item A transmission relation $\abr$ from $A$ to $B$ is {\it order-preserving}
  if links from $A$ to $B$ never cross.  That is, given $(a,b)$ and $(a',b')
  \in \abr$\,, it is never the case that $a<a'$ while $ b'<b$.  That means a
  symbol from $A$ to $B$ cannot be overtaken by another such symbol.  The
  transmission relation illustrated in Fig.\ \ref{fig:1}(b) is
  order-preserving; Fig.\ \ref{fig:X1} shows an instance of a transmission
  relation that does {\it not} preserve order: the link shown in red starts
  later than that shown in blue but arrives earlier.
\item A transmission relation from $A$ to $B$  is {\it sub-1-to-1} if no $A$-square is linked to more than one $B$-square and {\it vice versa}.
\item A transmission relation from $A$ to $B$ is {\it periodically timed} if
  $(\exists m,n) \text{ s.t. } (a,b)\!\!\in\,\abr\; \Rightarrow (a+m,b+n) \in
  \abr$\,. (Notes: (1) The case $n > m$ corresponds to $B$ stepping at a
  relative frequency $n/m$ with respect to $A$. (2) We say periodically timed
  rather than just periodic, because it is not required that the symbols be
  periodically distributed along the clock tapes, only that the links be
  periodic.)
\item
 Given a sub-1-to-1, order-preserving transmission relations from $A$ to $B$ and
 from $B$ to $A$, with $a,\;a' \in A$ and $b \in B$, if $(a,b)\in\, \abr$ and $(b,a')
\in \, \bar$\,, then we say the $b$ has an {\it echo count} relative to $A$ given
 by\vspace*{8pt} $a'-a$.  (This definition of echo count differs from that in
 \cite{14aop,16aop}.)
\end{enumerate}

\subsection{Networks of symbol-handling agents}
A set of agents and transmission relations among them specify a {\it
  network}.  Networks engender directed graphs (in the sense of vertices
connected by arrows).  Corresponding to (the clock tape of) an agent
is a graph consisting of a single path with a node for each square on the
clock tape and an arrow (directed edge, in graph terminology) from each node
to its successor (if it has one).  Two agents linked by a transmission
relation make a graph consisting of the two paths of the agents, linked by
arrows expressed by the transmission relation from one agent to the other.
Two agents and the two transmission relations back and forth between them
make a directed graph with arrows in both directions.  Similarly, networks
with more agents correspond to graphs that have pairs of agents and their
transmission relations as subgraphs.  Graphs representing networks have
non-intersecting paths for clock tapes of agents.

In the use of networks of symbol-handling agents to model various physical
and biological networks, a few concepts taken from graph theory, especially
Petri nets \cite{petri}, are helpful.
\begin{enumerate}
\item  Supposing a network of agents $A_\ell$ with $\ell$ in some index set of integers, the {\it forward reach of square $j$ of agent $\ell$} is
  \begin{equation}
    A_\ell(j)^{\bullet}  \coloneqq \{(A_k,m) | k\ne \ell \wedge  (j,m) \in 
    \,\stackrel{\xrightarrow{\hspace*{.7cm}}} {A_\ell A_k}\}.
  \end{equation}
  The red dots in Fig.\ \ref{fig:X2}(a) illustrate the forward reach of square $j$ of
  agent $A_1$.
\item Supposing a network of agents $A_\ell$ with $\ell$ in some index set of integers, the {\it backward reach of square $j$ of agent $\ell$} is
  \begin{equation}
    ^\bullet\!\! A_\ell(j) \coloneqq \{(A_k,m) | k\ne \ell \wedge  (m,j) \in 
    \,\stackrel{\xrightarrow{\hspace*{.7cm}}} {A_kA_\ell}\}.
  \end{equation}
  The red dots in Fig.\ \ref{fig:X2}(b) illustrate the backward reach of square $j$
  of agent $A_1$.
\begin{figure}[h]
    \includegraphics[height=2.5 in]{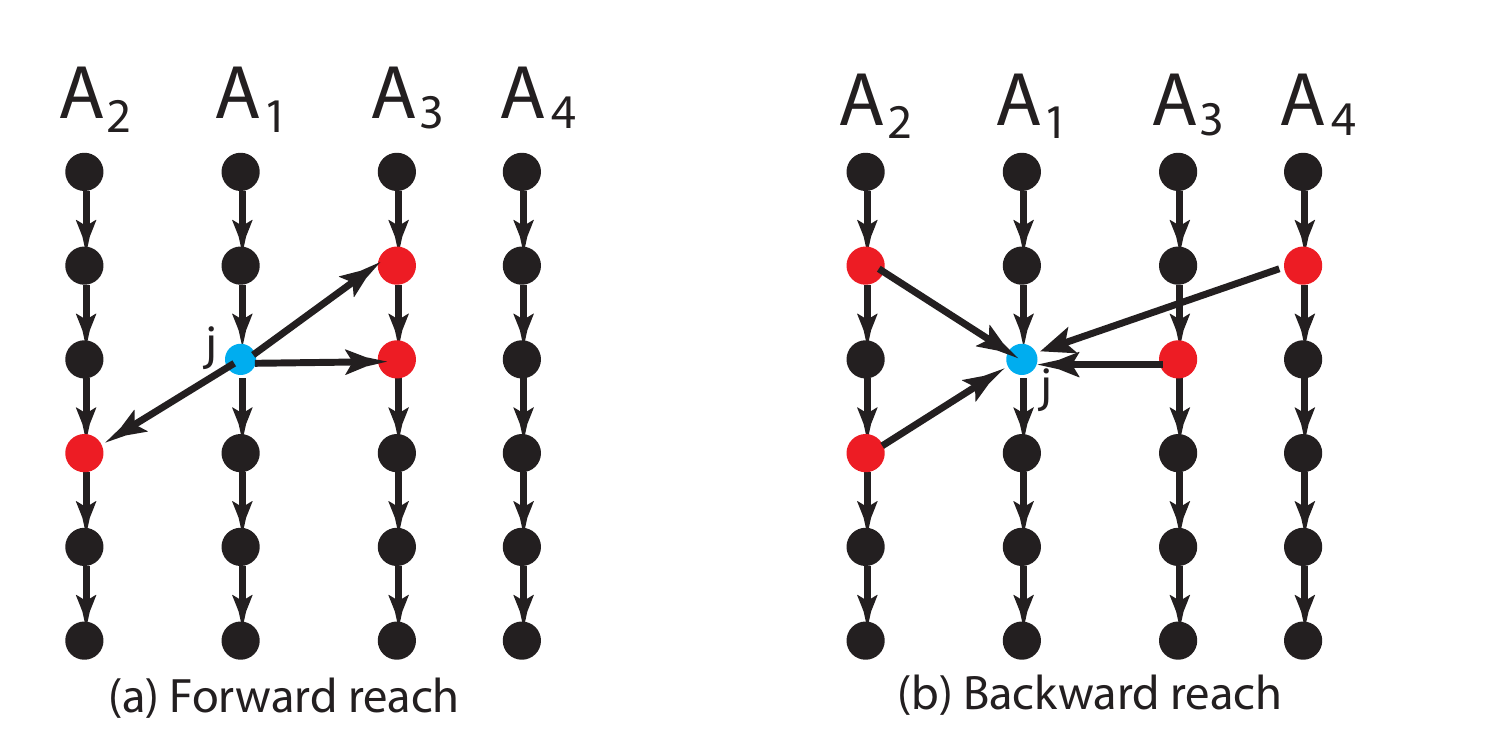}
    \caption{Forward and backward reach from \cite{JCS}.\label{fig:X2}}
  \end{figure}

\item $|A_\ell(j)^{\bullet}|$ denotes the number of clock-tape squares of other agents influenced by symbol $j$
  of agent $\ell$'s tape.  This is a measure of fan-out.  In Fig.\ \ref{fig:X2}(a) $|A_1^\bullet(j)|=3$.
  
\item $|^\bullet\!\! A_\ell(j)|$ denotes the number of symbols arriving at square $j$
  of the clock tape of $A_\ell$. This is a measure of fan-in. In Fig.\ \ref{fig:X2}(b)
  $|^\bullet\!\! A_1(j)|=4$.
\item A network is sub-1-to-1 if
  \begin{equation}
    (\forall\; \ell, j) |A_\ell(j)^{\bullet}|\le 1 \text{ and }\; |^\bullet\!\! A_\ell(j)|\le 1. 
  \end{equation}
\item A network without closed circuits of arrows expresses a partial order
  which allows one to speak of ``later'' and of ``concurrent''\cite{petri}.  A
  square $B_k(i)$ is {\it later} than a square $A_\ell(j)$ if there is a path
  of arrows from $A_\ell(j)$ to $B_k(i)$.  Two squares for which there is no
  such path from one to the other are {\em concurrent}.

  Networks (based on the clock tapes of agents) that are not partial orders
  are {\it acausal} in the sense that a later symbol can influence the
  writing of an earlier symbol.  All applications that we so far envisage for
  networks rule out acausal networks.
\end{enumerate}

The case $|A_\ell(j)^{\bullet}| > 1$ (forward reach greater than 1)
corresponds to broadcasting by $A_\ell$ of the symbol on square $j$.  The
case $|^\bullet\!\! A_\ell(j)| >1$ (backward reach greater than 1)
corresponds to the writing of symbol on a square $j$ of the clock tape of
$A_\ell$ being influenced by more than 1 symbol arriving during period $j$.
(Think of listening to a symphony.)

\subsection{Picturing a population of agents}
In evolutionary biology, one considers populations of organisms that are born and that die.  Symbolic communications among agents representing
organisms of such a population involve no fixed network, but instead involve
the entrance and termination of agents with their clock tapes, leading to a
dynamically evolving network.  Viewing such a dynamic network in terms of the
clock-tape records, we can portray the entrance and termination of agents as
in Fig.\ \ref{fig:4}:
\begin{figure}[h]
  \hspace*{1 in}
    \includegraphics[height=4 in]{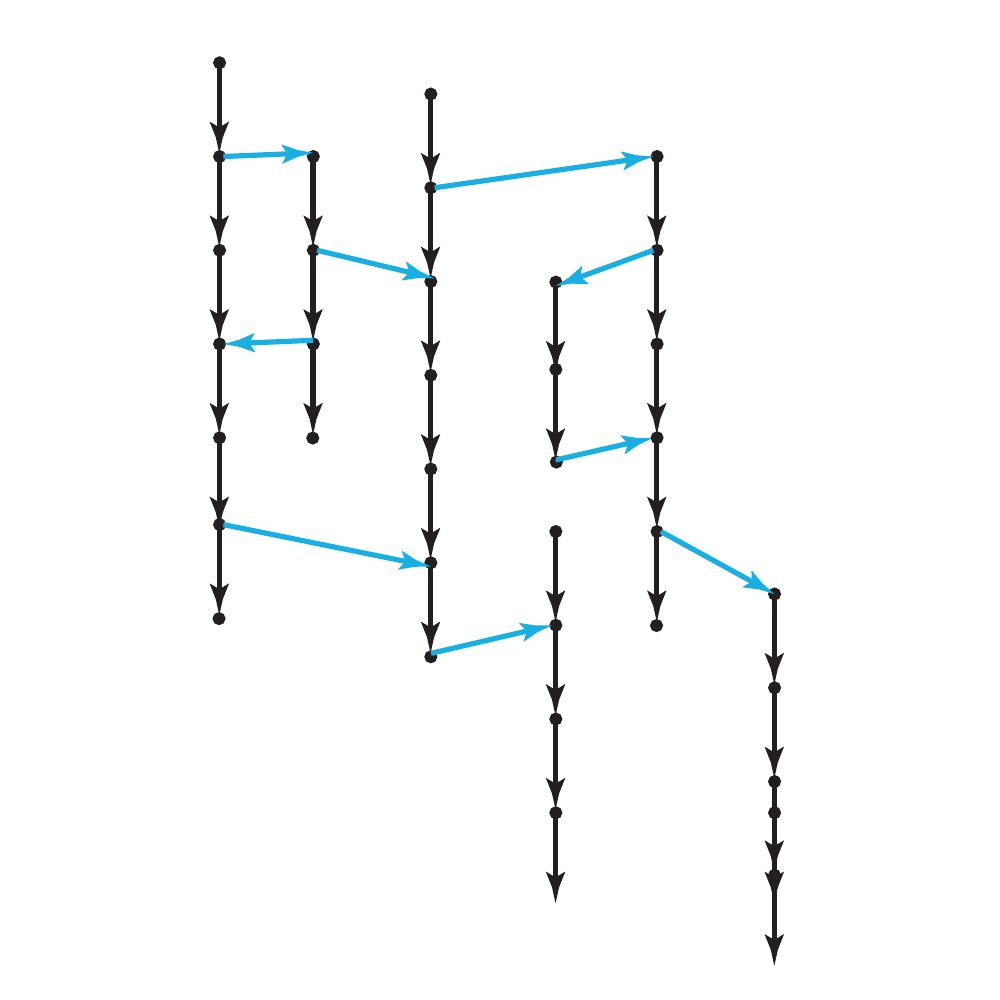}
    \caption{Showing births and deaths in clock tapes of a population of
      symbol-handling agents from \cite{JCS,19spie}.\label{fig:4}}
  \end{figure}

\subsection{Cycles}
\begin{figure}[h]
  \hspace*{1 in}
    \includegraphics[height=4 in]{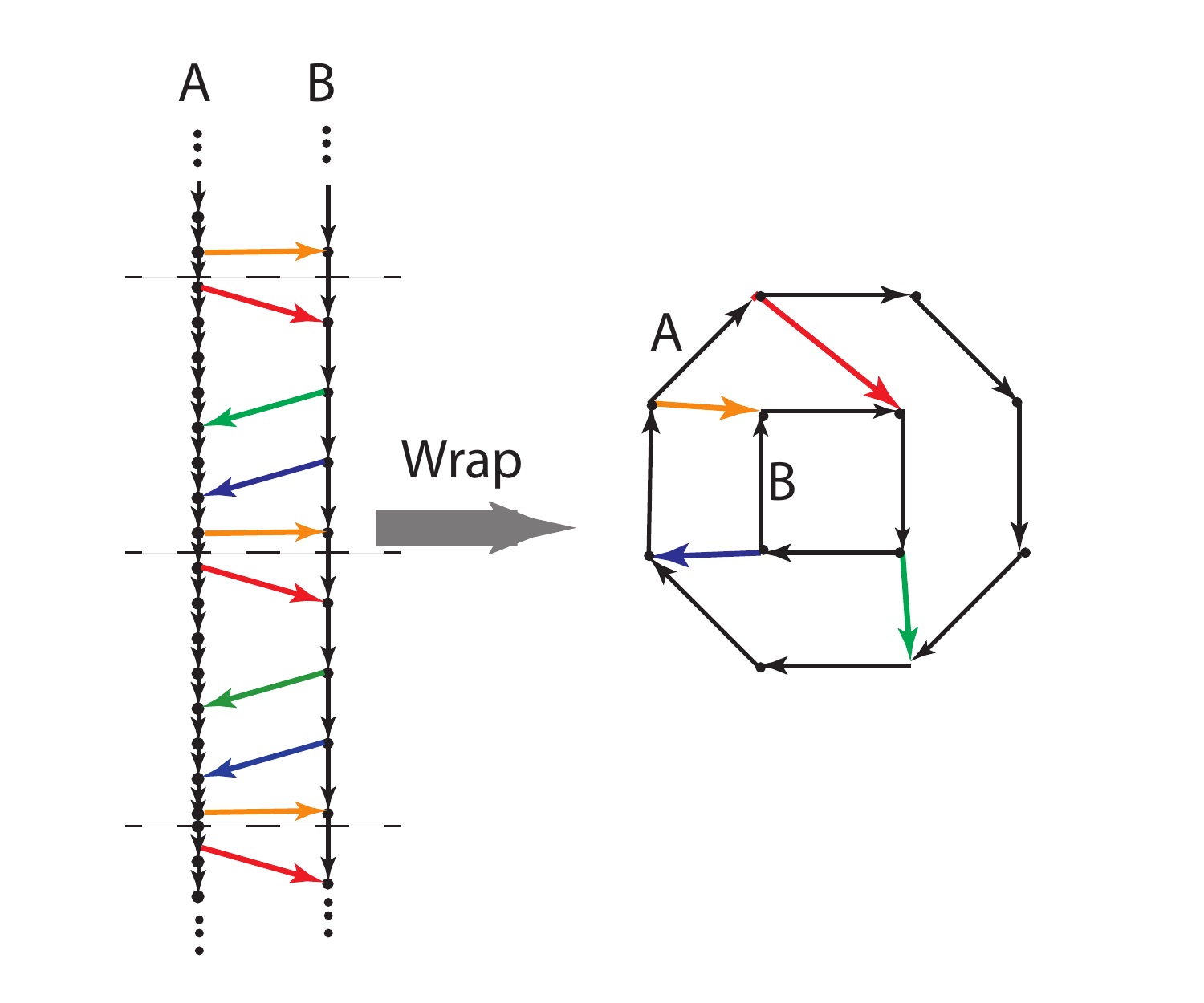}
    \caption{Wrapping a periodically timed network into a cyclic graph from \cite{JCS,19spie}: $A$
      wraps to octagon; $B$ wraps to square.\label{fig:3}}
\end{figure}

Sometimes it is desirable to emphasize the cyclical nature of agents,
whether as investigators or as subjects of investigation, or both.  For
example, Sec.\ \ref{sec:5} recounts an investigation of memory elements,
involving a sequence of trials, each of which proceeds through periodically
timed phases.  To portray the cyclical aspect of such a case, one can ignore
any non-periodic transmissions, thereby arriving at a periodic graph (or
several disconnected periodic graphs.)  A periodic graph can be
wound into a cyclic graph, as shown in Fig.\ \ref{fig:3}.  Winding wraps a
repeating stretch of the periodic partial order into a graph in which each
agent is mapped into a cycle.

In contrast to a partial order, in a cyclic graph, there is no two-place
{\it before} relation, as in ``$a$ before $b$.'' Instead one has a
three-place {\it between} relation, e.g ``$b$ between $a$ and $c$.''  If the
period of a graph wound into a cyclic graph corresponds to three or more
squares of every agent, the winding is called {\it loop-free}\cite{cyclic}.
A {\it loop-free} winding of a periodic partial order generates a cyclic
partial order---a structure with a myriad of interesting features
\cite{cyclic}.  (If the winding is not loop-free, there are too few nodes on
some cycle to admit the relation of ``between''.)  Haar points out depth of
the mathematics concerning windings of partially ordered sets \cite{cyclic}.

\subsection{Avenues of application of the theory of symbol-handling agents}
We turn now from the mathematics of the theory of symbol handling agents to
examples of avenues of application of that theory.  Transmission relations on
clock tapes offer an underpinning to the theory of spacetime, as well as
opening up alternatives to spacetime.  In the theory of relativity, a
mathematical spacetime is a set of events, and one associates an event of a
spacetime with a physical ``event'', e.g.\ a flash of light.  In place of
``events'' one can think of records of symbols on squares of clock tapes
linked by transmission relations.  This freedom allows the exploration of
situations in which times and places can be constructed to suit particular
situations.  The theory of symbol-handling agents based on transmission
relations offers templates for the design of experiments.  Here are three
examples:
\begin{enumerate}
\item  Attaching locations to events is a basic activity of physics, typically
done by assigning spacetime coordinates, as articulated in 2000
resolutions of the International Astronomical Union:
\begin{quote}
The underlying concept in relativistic modeling of astronomical observations
is a relativistic four-dimensional reference system. By reference system, we
mean a purely mathematical construction (a chart or a coordinate system)
giving ``names'' to spacetime events \cite{soffel03}.
\end{quote}
As a reference system, the IAU resolutions assume a curved spacetime with a
metric tensor field chosen to represent the exterior of the Earth.  Thus names of
events depend on the assumption of a curved spacetime with a particular
choice of metric tensor field.  An alternative of theoretical, and in some
cases practical, interest is to name events by a clock as the place at which
the event occurred and the reading of that clock as the (local) ``time'' of
the event.  This alternative makes the names correspond to actual or
imaginable measured data, leaving one free to consider how the data might
suggest differences from the IAU metric tensor in alternative proposals for Earth's gravitation.
\item
Undersea acoustic networks are of interest for investigating the behavior of
cetaceans (e.g. porpoises) that communicate using sound.  It
appears interesting to use the clock-tape perspective to construct
times and places based on sonar communications in rhythms adapted to the
communications of the cetaceans.
\item
Animal nervous systems function in a variety of rhythms, and, we suspect,
involve the manipulation of symbols.  As it develops before and after its
birth, an animal develops its own system of times and places.  The freedom
for an investigator to make and to test hypothesis that adapt times and
places for stimulating the animal nervous system to the animals own
development of rhythms looks promising.
\end{enumerate}

\section{Logical synchronization: how distinctions need gradations}\label{sec:4}
Attention to the necessities of implementation is fostered by the statement,
made in the introduction, of a certain independence of the workbench of
experiment from any theory.  In this section we introduce a kind of
synchronization quite different from Einstein's, arising from behavior found
in actual digital systems that implement the transmission relations
discussed theoretically in the preceding section.

Unlike the imagined proper clock of relativity theory, a physical clock
oscillates through phases---think of a swinging pendulum.  Without the
phasing there would be no ``ticks'' to count.  Special relativity, however,
is based on an abstraction that makes ticks invisible.  Early in
the paper in which he introduced special relativity, Einstein asserts that
judgments in which time plays a role are judgments of the coinciding of events.
That assertion comes with an asterisk pointing to an interesting footnote [in our translation from the German]:
\begin{quote}
  The inexactitude that lurks in the concept of the coinciding of two events
  at (approximately) the same place has be skated over by an abstraction that
  we leave undiscussed \cite{einstein05}.
\end{quote}
For a theory of symbol handling, this abstraction obscures the a critical
issue.  An agent, like a digital computer, is stepped by a clock that cycles
through periodic phases.  If writing a symbol into memory overlapped
temporally with reading a symbol from memory, the result would be confusion
of logic. In order to avoid this confusion, a symbol transmitted to an agent
must arrive only during a particular phase of the receiving agent's clock and
not during other phases \cite{16aop}. This constraint, which we call {\it
  logical synchronization}, requires leeway in the arrival time; one can't
ask for a point coincidence.  In contrast to Einstein synchronization, the
concept of logical synchronization has this leeway built into it.  The need
for logical synchronization, long known to engineers of digital
communications \cite{meyr}, is reminiscent of a game of catch, in which a
player cycles through phases of throwing and catching a ball, or more simply,
a spoken dialog in which each person alternates between speaking and
listening.

\subsection{Logical synchronization vs. Einstein synchronization}
As discussed in
\cite{14aop,16aop}, logical synchronization has both freedoms and
constraints relative to Einstein synchronization.  Freedoms include:
\begin{enumerate}
\item Unlike Einstein synchronization, clock readings at transmissions and
  receptions are allowed a certain leeway.
\item Unlike Einstein synchronization, the logically synchronized clocks can
  differ in frequency.  That is because the conditions for logical
  synchronization are not required for all periods of the clock curves, but
  only for those periods linked by the transmission of a symbol \cite{14aop}.
\item Because of the freedom to vary clock rate relative to a proper clock,
  two agents in relative motion in a flat spacetime can maintain
  logical synchronization, even though Doppler shift precludes Einstein synchronization.
\end{enumerate}

Constraints include:
\begin{enumerate}
 \item Transmissions and receptions are restricted to appropriate clock phases.
 \item Consider several agents thought of as in a spacetime, communicating
   symbols carried as light pulses.  The requirement of logical
   synchronization strongly constrains the possible transmission relations.
   This constraint is discussed in \cite{14aop} as the ``stripes in
   spacetime'' imposed by logical synchronization; it corresponds to ``you
   can't synchronize with everybody at once, so you have to make choices''.
\end{enumerate}

\subsection{Extra-logical clock adjustment to maintain logical synchronization}
In many situations, to maintain the arrival of symbols within the leeway
allowed by logical synchronization, agents must more-or-less continually
adjust the tick-rates of their clocks.  The adjustments of clock rates
necessary to the maintenance of logical synchronization are steered by a
feedback loop that estimates phase deviations from the aiming point.  To
sense deviations within the leeway, an agent must reach beyond logical
operations on symbols, for the simple reason that the logic of symbol
handling has to be oblivious to those deviations.
\begin{prop}\label{prop:X}
The timing of symbol arrival within the allowed phase cannot be registered by
the process that recognizes distinct symbols.
\end{prop}

\noindent{\it Proof}: the recognition of a symbol depends on indifference to
the timing of arrival within the allowed \vspace*{8pt} leeway.

It follows that distinction-bearing symbols can't be the whole story, for
they cannot function without agents attending to gradations.  Thus auxiliary
mechanisms are necessary to supply an agent with information to guide the
steering of its clock rate.  Steering of clock rates so as to maintain
logical synchronization is often automated to function according to an
algorithm that responds to graded deviations of the phases of arriving
symbols registered over some running number of cycles.  The computational
complexity of the algorithm is in many cases minimal.  But because, even in
principle, deviations are unpredictable \cite{14aop}, no algorithm, no matter
how complex, can anticipate deviations so perfectly as to eliminate them.
Choosing an algorithm to steer clocks requires that an agent reach beyond
logic to make a guess.

In the next Section we go into behavior on the work bench that illuminates
the gradations necessary to dealing with distinct symbols.

\section{When the coin lands on edge}\label{sec:5}
In Sec.\ \ref{sec:3} we mostly focused on theoretical transmission relations
on the clock tapes of agents, relations that can be written to sit still on
the blackboard.  In Sec.\ \ref{sec:4} we enriched the theory of symbol
handling by considering the need for logical synchronization, essential to
implementing designs on the workbench based on theoretical transmission
relations.  But logical synchronization does not just happen; agents must
maintain it by steering clock rates.  The steering of clock rates is dynamic,
involving not only distinctions but also indistinct arrivals of symbols
within a phase.  Logical synchronization depends on agents attending to
graded transitions between distinctions.  Here we discuss the gradations that
have to be dealt with in order to implement logical distinctions on the
workbench.

We start with the question: what happens when the arrival of a symbol fails
to meet the conditions of logical synchronization?  We show how an agent's
act of receiving a symbol outside a receptive phase is like flipping a coin
that lands on edge, resulting in logical confusion, sometimes referred to as
a `glitch'.  We go beyond our earlier discussions \cite{05aop,16aop}, by
relating the glitch to evidence of logical confusion pictured on clock tapes.

Logic on the workbench is built from physical {\it NAND gates} used to
construct a digital computer.  On the blackboard, a NAND gate is thought of
as implementing the NEGATION of the Boolean function ``AND'', but a
NAND gate on the workbench moves.  It has two input wires and an output wire;
on all three wires, voltages implementing Boolean values 0 or 1 undergo
changes.  When voltages are held constant for a little while on its input wires
the NAND gate generates, after a delay, a voltage on its output wire---a high
voltage for a 1 unless both input wires have high voltages, in which case the
output is a low voltage for 0.  The phrase ``after a delay'' is one hint that
logic on the workbench differs from blackboard logic.  A digital system,
composed of NAND gates must be temporally organized, which requires that
some of the inputs of its NAND gates are driven by clocks.  Only then can
the digital system deal coherently with changes in inputs and outputs.

\begin{figure}
    \includegraphics[height=4.5 in]{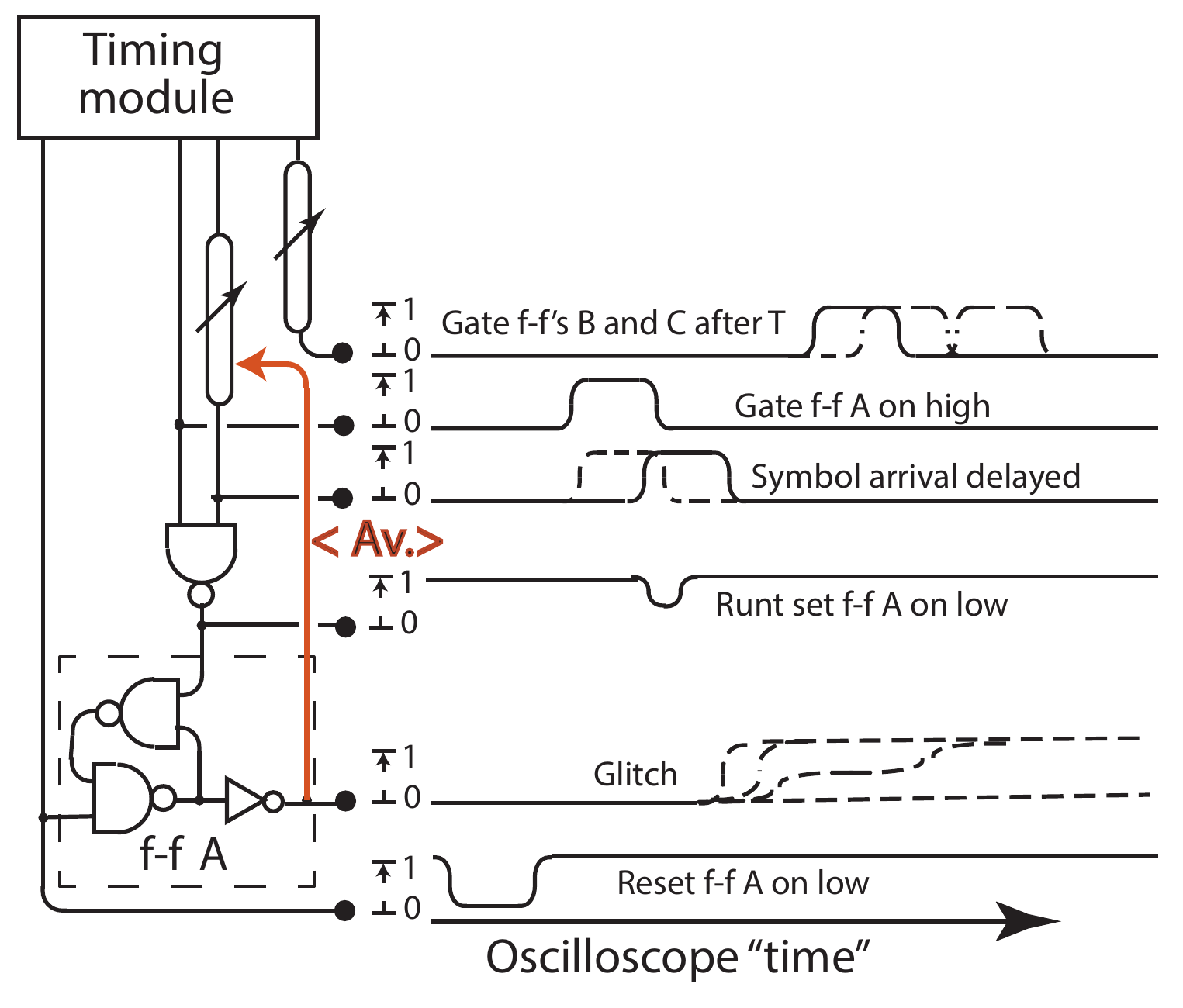}
    \caption{Oscilloscope traces of flip-flop $A$ under test.\label{fig:osc}}
  \end{figure}

A pair of cross-coupled NAND gates called a {\it flip-flop} implements a
square of a clock tape on which can be written a single bit as a 0 or a 1.
The two NAND gates of a flip-flop form an unstable balancing device, the
electronic analog of a hinge that records a 1 if flipped one way or a 0 if
flopped the other way.  A gated flip-flop is a flip-flop with one input
preceded by a third NAND gate that acts as a valve.  If open, it allows
the ``hinge'' to flip or flop, and if closed, prevents the ``hinge'' from
flipping or flopping.  Fig.\ \ref{fig:osc} shows how two NAND gates form a
flip-flop. It also shows the NAND gate that precedes the flip-flop and acts
as a valve.  The cross coupled NAND gates of a flip-flop feedback on
themselves, thereby providing another hint of a difference between
blackboard logic and bench logic.

As discussed in Sec.\ \ref{sec:4}, the clocks of digital systems step
flip-flops through phases of a cycle, so there is a phase for changing the
symbols on the inputs and a distinct other phase during which symbols appear
on the outputs.  Translated into terms that relate to an experiment on
flip-flops, our question becomes: what happens if a symbol arrives just as an
agent's clock closes the valve on the phase in which an agent's flip-flop can
accept a symbol?
 
A trial of the experiment begins by resetting the flip-flop to 0.  When the
flip-flop has been reset to 0, a 1 arriving during a phase in which the valve
is open flips the hinge over, so that the flip-flop generates an output of 1.
So, to repeat our question: what happens when the symbol 1 arrives not while
the valve is open, but just as the valve is closing?  One might guess that
what happens is random, i.e., either the flip-flop generates an output of 1
or it generates an output of 0, but that guess is, at best, misleading.

\begin{figure}[h]
  \hspace*{1 in}
    \includegraphics[height=3.75 in]{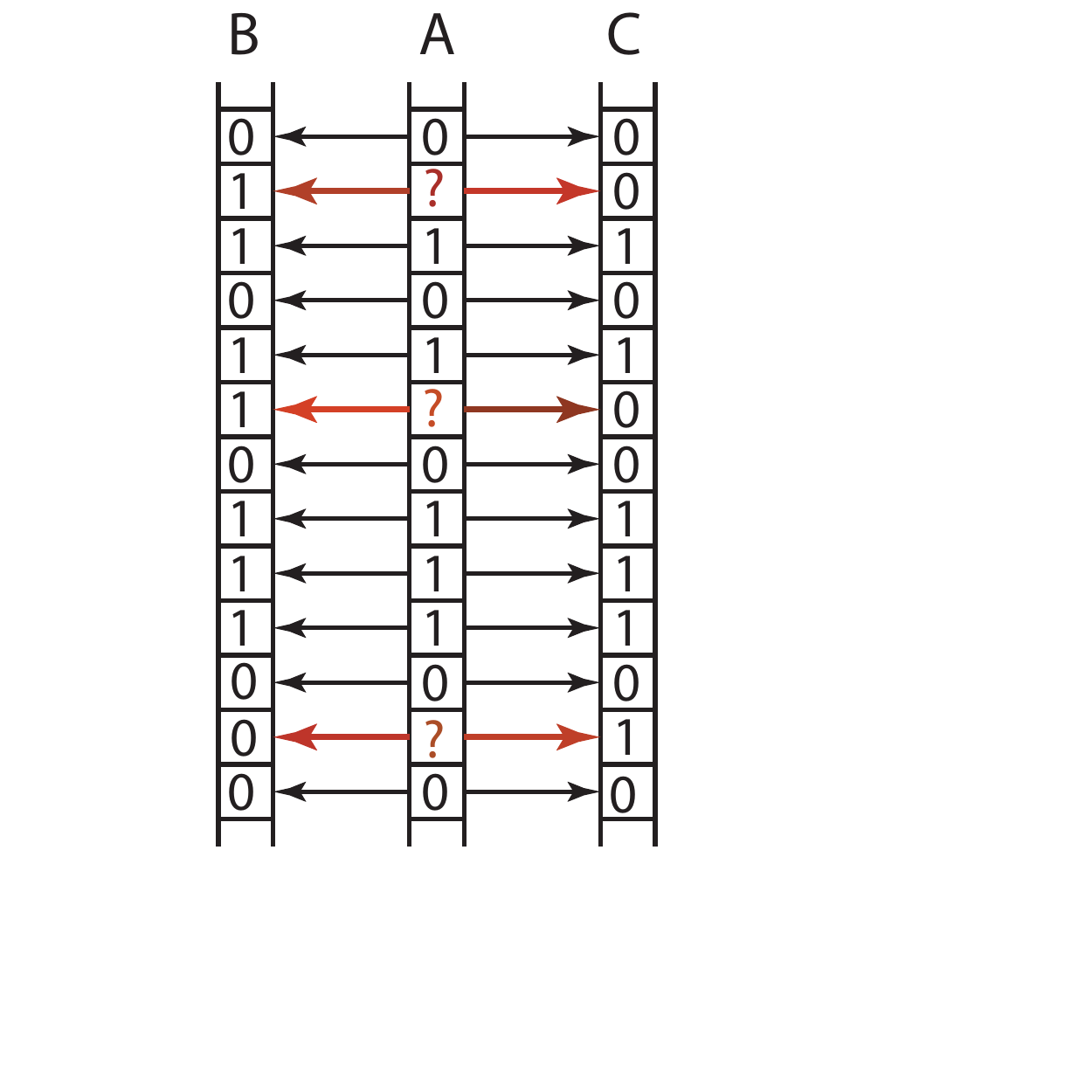}\vspace*{-1 in}
    \caption{Disagreements between Agents $B$ and $C$ viewing $A$.\label{fig:glitch1}}
  \end{figure}

We and others (e.g. \cite{scfl}) have experimented to find out what happens.
Our focus on symbols led us to counting evidence of glitches expressible by
relations among clock tapes.  We arrange a clock, shown as the ``Timing
module'' in Fig.\ref{fig:osc}, to drive a sequence of trials of a flip-flop
$A$ that, after a variable delay $T$, is viewed by a matched pair of
flip-flops $B$ and $C$.  Viewing the flip-flops, including their clocking, as
agents, we can display the experimental results on clock tapes for $A$, $B$,
and $C$.  Successive squares on the clock tapes of $A$, $B$, and $C$ are
generated in lock-step, one each per cycle of the clock that drives the
trials.  Fig.\ \ref{fig:glitch1} shows the form of evidence, in which the
evidence of glitch is seen when a $B$-square and a $C$-square linked to a
given
$A$-square disagree: one shows a 0 while the other shows a 1.  In this form
of evidence, any given square of a clock tape holds a distinct symbol and not
any other symbol; however, to show how the experiment works, one has to show
what goes on within phases of the clock cycle.

The left margin of Fig.\ \ref{fig:osc} shows the electrical circuit diagram
of flip-flop $A$ under test, with the probe points indicated by {\Large
  \textbullet}.  The circuit shows the three NAND gates mentioned above,
along with two variable delays.  On the right of the figure are oscilloscope
traces of voltage vs. time for the probe points, as told by the sweep of the
electron beam of the oscilloscope.  The traces illustrate the effect of a
symbol arriving just as the gate for its acceptance is closing.  The
immediate effect is the generation of a runt pulse\,---\,enough of a pulse to
perturb flip-flop $A$, but not enough to necessarily set it.  The result is
various.  In some trials, flip-flop $A$ is set promptly; in other trials it
is not set at all, and, in the more interesting trials, flip-flop $A$ is put
into an unstable equilibrium, where it teeters, indefinitely, before
generating a definite high or low output.
\begin{prop}
  In some cases in which a symbol arriving at a clocked flip-flop $A$ comes
  late or early relative to the receptive phase determined by the clock, the
  flip-flop teeters indefinitely, before, eventually ``falling over'' to
  generate a distinct low or a high voltage.  When $A$ transmits its output
  to the inputs of two flip-flops $B$ and $C$, these can disagree about $A$'s
  state; the likelihood of disagreement diminishes with the waiting time $T$
  prior to $B$ and $C$ being gated to read $A$'s output.
\end{prop}

Our experiment required a little invention to make symbol arrivals straddle a
boundary between phases (just the opposite of steering clock rates so as to
maintain logical synchronization). We needed to shape the runt pulse so as to
put flip-flop $A$ into the teetering condition often enough to study it.  The
runt pulse is shaped by the difference in two delays, and the delays drift
unpredictably with temperature variations and other unknown influences.  When
we set the delays statically, the drift made $A$ register a sequence of all
1's or a sequence of all 0's, with no glitches.  To counter the effect of
drifts we fed back a short-term running average of $A$'s outcomes to control
one (voltage variable) delay line.  While trying to adjust the delays
statically from the behavior of the oscilloscope traces failed, the use of
feedback worked fine.

In summary, to avoid logical inconsistencies (``$B \ne C$''), agents
receiving symbols must ``avoid looking'' during transitions between symbols .
This ``avoiding looking'' during a transition from between logical
values---i.e.\ logical synchronization---is no passive circumstance, but
generally requires agents to actively steer the tick rates of their clocks.
Furthermore, and this we emphasize, maintaining logical synchronization
requires agents to attend to the intrinsically unpredictable gradations.  The
attempt to impose distinctions on these gradations would put a flip-flop into
an unstable state.
\begin{prop}\label{prop:influence}
The order of events asserted by a flip-flop in a race condition can be
decided by extraneous influences, no matter how small: the teetering of a
flip-flop allows, for a little while, the future to affect the record of the
past.
\end{prop}
\begin{prop}
  The registering of disagreements under fan-out proved a superior
  measurement technique for recording borderline cases.
\end{prop}

\subsection{Extended lessons from the glitch}
\noindent Now we introduce the following\newline
\noindent{\bf Assumption}
({\bf Principle of the balance}) {\it The measurement of the temporal order of
symbol arrivals requires balancing one arrival against a single other
arrival, e.g.\ by use of a flip-flop, making measured order of arrival a
binary relation.}

From this principle and Prop.\ \ref{prop:influence} it follows that
\begin{prop}\label{prop:5_4}
  The decision of a close race is necessarily beyond logic.
\end{prop}
The principle of the balance also implies 
\begin{prop}\label{5_6}
  Measuring the order of arrival in an $n$-way race requires measuring
  pairwise, that is, measuring the order of arrival of $n(n-1)/2$ two-way
  races.
\end{prop}

Describing a race in terms of theoretical ``arrival times'' encounters a
conflict with the workbench.  On the blackboard, a time is expressed as a
real number $t$ and a reference \cite[Sec 1.1 1.1]{vim}.  Then for any two
such blackboard ``times'' $t_1$ and $t_2$ there are the three mathematical
possibilities.  $t_1 < t_2$, $t_1=t_2$, or $t_1 > t_2$.  On the workbench, in
the absence of logical synchronization, the glitch tells us that ``='' is
unstable.
\begin{prop}\label{prop:5_7}
  Indeterminacy in two-way races implies possible non-transitivity in
  races among three or more; e.g. in a three-way race of $a$, $b$, and $c$,
  there can be cases of finding $a<b$, $b<c$, with $c<a$.
\end{prop}

\begin{prop}\label{prop:5_3}
The use of real numbers on the blackboard to express timing under
  race conditions conflicts with experimental evidence.
\end{prop}
\vspace*{-.1 in}
\section{Concluding remarks}\label{sec:6}
``Mathematics is based on the idea of a distinction'' \cite{kauffman}, and
the conveying of distinctions by the use of symbols starts with life itself,
e.g. in the bases of DNA.  The symbols expressing formulas of mathematical
logic can sit still on the blackboard, but logic on the workbench is the
logic of devices that move.  Motion creates a problem on the workbench
foreign to logic on the blackboard, namely the deciding of temporal order:
which symbol came before the other?  We have tried to let the device on the
workbench that decides temporal order---the flip-flop, with its
teetering---tell its story, a story that binds the communication of
distinctions to logical synchronization with its dependence on the
unpredictable teetering, richer than can be captured by ``measurement
uncertainty.''  Three more remarks:
\begin{enumerate}
\item With the recognition of symbol-handling as part of physics and part of
  life, the role of clocks reaches beyond ``telling time'' to the opening and
  closing of gates necessary to the coherent communicating of
  distinctions.
\item Without logical synchronization, agreement about distinctions
  is impossible.
\item With logical synchronization, the arrival of a symbol, as recorded on a
  square of a clock tape, is {\it objective} in the sense that one expects
  that two agents to which the square fans out will agree on the symbol.
  Objectivity in this sense endures after we give up any aspiration to final
  ``truth,'' as we must in light of the incompleteness theorem discussed in
  \cite{18incomplete}.
\end{enumerate}

The work reported here opens a door to dealing with the timing of symbolic communication in a way that supplies a previously unavailable underpinning to concepts and implementations of ``times and places''.  There is a lot more to explore.  We have discussed the maintenance of logical synchronization, once that synchronization is in place.  Left to the future is the challenging topic of two agents that seek to acquire logical synchronization so that they can communicate.
From the engineering world, we can point to the negative result that there can be no deterministic upper bound on how many cycles that acquisition may require
\cite{meyr}.

\section*{Acknowledgments}
It is our pleasure to acknowledge helpful conversations with Kenneth Augustyn,
Paul Benioff, and John Realpe-G\'omez.  We are also greatly indebted to anonymous reviewers for suggesting important improvements.

\appendix
\section{Synchronization and the theory of general relativity}
For the proper clocks of the theory of special relativity, Einstein defined a
form of synchronization, {\it Einstein synchronization}, that permeates
theoretical physics.  For example in the International System of Units (SI),
Einstein synchronization guides the definitions of the measuring units for
time and for length.  
\subsection{The meter in relation to clock readings as defined in special relativity}
The SI meter is ``the length of the path traveled by light in vacuum during a
time interval of 1/299 792 458 of a second'' \cite{brochure9}.  This `time
interval' rests on the concept of an inertial frame of special relativity.
Every clock fixed to an inertial frame is Einstein synchronized to every other
such clock. Einstein synchronization relates readings of one clock to readings of
another clock.  Let $t_A$ be a reading of clock $A$ at the emission of a
light pulse that reaches $B$ at $t_B$, and $t'_A$ the reading of $A$ at the
receipt of an echo reflected from $B$ at $t_B$.  Looking at $t_A$ and $t'_A$
as functions of $t_B$, clock $B$ is Einstein synchronized to clock $A$,
relative to the inertial frame, provided that, for all $t_B$,
\begin{equation}\label{eq:esyn}
  t'_A -t_B = t_B-t_A.
\end{equation}
\begin{prop}\label{prop:2_0}
      The time interval in the definition of the meter denotes the
      difference between clock readings of Einstein-synchronized proper
      clocks at the two ends of a light path.
    \end{prop}
For $c$ the speed of light, the SI length of a path from $A$ to $B$ is
$c(t_B-t_A)$, and thus invokes readings of separated, Einstein-synchronized
proper clocks.  Note that even in special relativity, Doppler shift precludes
Einstein synchronization of proper clocks moving relative to each other.

\subsection{Einstein synchronization drastically restricted by spacetime curvature}
Spacetime curvature changes the story.  It is known that in a generic curved
spacetime of the theory of general relativity, no grid of exactly
Einstein-synchronized proper clocks is possible. Because curved spacetimes
are locally flat, deviations from Einstein synchronization are often small;
however, the astounding stability of today's optical atomic clocks makes
small deviations from synchronization measurable and of physical interest, as
in the detection of gravitational effects.  For a second example, Coordinated
Universal Time (UTC) is distributed by clocks that, even in theory, require
their tick rates to be adjusted to compensate for gravitation.

\subsection{Clocks as expressed in general relativity}
In the theory of special relativity a clock fixed to an inertial frame is
expressed by a straight, timelike line.  Turning from the flat spacetime of
special relativity to the curved spacetimes of general relativity, one
expresses a clock in terms of a timelike curve in a manifold \cite{07perlick}:
\begin{quote}
      Here and in the following, our terminology is as follows. A
      general-relativistic spacetime is a 4-dimensional manifold $M$ with a
      smooth metric tensor field $g$ of Lorentzian signature and a time
      orientation; the latter means that a globally consistent distinction
      between future and past has been made. A clock is a smooth embedding
      $\gamma \colon t \rightarrow \gamma(t)$ from a real interval into $M$
      such that the tangent vector $\dot{\gamma}(t)$ is everywhere timelike
      with respect to $g$ and future-pointing. This terminology is justified
      because we can interpret the value of the parameter $t$ as the reading
      of a clock. Note that our definition of a clock does not demand that
      ``its ticking be uniform'' in any sense. Only smoothness and
      monotonicity is required \cite{07perlick}.
 \end{quote}
 We will speak of reparameterization of the embedding that specifies
 a clock as ``an adjustment of the tick rate of the clock''.
 
Instead of an inertial frame, for a curved spacetime one has
an ``observer field'':
\begin{quote}
     By an observer field on a general-relativistic spacetime we mean a
     smooth vector field $V$ which is everywhere timelike and
     future-pointing. An observer field $V$ is called a standard observer field
     if $g(V, V ) = −1$. According to our earlier terminology, integral curves
     of observer fields are clocks, and integral curves of standard observer
     fields are standard clocks with the usual choice of time unit. For the
     sake of brevity, we will refer to the integral curves of an observer
     field $V$ as to “clocks in $V$ ”. Note that $V$ fixes the parametrization for
     each of its integral curves uniquely up to an additive constant, i.e.,
     for each clock in $V$ there is still the freedom of “choosing the zero
     point on the clock’s dial” \cite{07perlick}.
 \end{quote}

For a generic curved spacetime, we can say something about the issue of
trying to Einstein-synchronize clocks in a {\it radar neighborhood}, which is
a neighborhood too large to be considered flat, but ``small'' enough to avoid
extreme gravitational effects \cite{07perlick}.  More precisely, given clocks
$A$ and $B$ within a radar neighborhood, for an event $b\in B$ there is
precisely one light ray from $A$ to $b$, and one light ray from $b$ to $A$.

Although no inertial frame of Einstein-synchronized proper clocks is
possible in a curved spacetime, there exists adjustments of the tick rates of
selected pairs of clocks that can make them Einstein synchronized.
\begin{prop}\label{prop:2_2}
For any two non-intersecting clocks following given timelike trajectories
within a radar neighborhood of a generic curved spacetime, there exist tick
rates, in general varying, for which the two ``improper'' clocks can be
Einstein-synchronized.
\end{prop}
For a flat spacetime, the needed  adjustment of (possibly moving) clocks is illustrated in Fig. 4 of \cite{14aop}, and the same procedure works in a radar neighborhood of a curved spacetime.  However, when more than two clocks are considered in a curved spacetime, it is in general impossible to Einstein synchronize each clock to all the others.

A ``radar distance'' can be defined for improper clocks in a curved spacetime,
analogous to distance as defined in special relativity, but in a curved spacetime,
radar distance is neither transitive nor symmetric \cite{07perlick}.  

\begin{prop}\label{prop:2_3}
  Assuming a generic curved spacetime, as the maximum radar-distance across a
  network of more than two clocks increases, the minimum possible deviations
  from Einstein synchronization also increase, even when adjustable clocks
  are allowed.
\end{prop}
From Prop.\ \ref{prop:2_3} we arrive at Prop.\ \ref{prop:2_4} of Sec.\ \ref{sec:2}.

\end{document}